\pdfoutput=1
\documentclass[aps,prb,twocolumn,amsfonts,amssymb,amsmath,floats,floatfix,showpacs,preprintnumbers,superscriptaddress]{revtex4-1}
\usepackage{amsmath,amssymb}
\usepackage{graphicx}
\usepackage{color}
\usepackage{hyperref}
\usepackage{bm}
\usepackage{setspace}
\usepackage{appendix}

\hypersetup{
colorlinks=true,
urlcolor= blue,
citecolor=blue,
linkcolor= blue,
bookmarks=true,
bookmarksopen=false,
}

\newcommand{\Nd}{NdFeAsO$_{1-x}$F$_x$}
\renewcommand{\vec}[1]{\bm{#1}}

\begin{document}

\title{Single Vortex Pinning and Penetration Depth in Superconducting NdFeAsO$_{1-x}$F$_x$}

\author{Jessie T.\ Zhang}
\thanks{The first three authors made contributions of equal importance.}
\affiliation{Department of Physics, Massachusetts Institute of Technology, Cambridge, MA 02139, USA}
\author{Jeehoon Kim}
\thanks{The first three authors made contributions of equal importance.}
\affiliation{Department of Physics, Harvard University, Cambridge, MA 02138, USA}
\author{Magdalena Huefner}
\thanks{The first three authors made contributions of equal importance.}
\affiliation{Department of Physics, Harvard University, Cambridge, MA 02138, USA}
\author{Cun Ye}
\affiliation{Department of Physics, Tsinghua University, Haidian, Beijing 100084, China}
\author{Stella Kim}
\affiliation{Ames Laboratory, U.S. DOE and Department of Physics and Astronomy, Iowa State University, Ames, Iowa 50011, USA}
\author{Paul C.\ Canfield}
\affiliation{Ames Laboratory, U.S. DOE and Department of Physics and Astronomy, Iowa State University, Ames, Iowa 50011, USA}
\author{Ruslan Prozorov}
\affiliation{Ames Laboratory, U.S. DOE and Department of Physics and Astronomy, Iowa State University, Ames, Iowa 50011, USA}
\author{Ophir M.\ Auslaender}
\affiliation{Department of Physics, Technion -- Israel Institute of Technology, Haifa 32000, Israel}
\author{Jennifer E.\ Hoffman}
\email[]{jhoffman@physics.harvard.edu}
\affiliation{Department of Physics, Harvard University, Cambridge, MA 02138, USA}
\date{\today}

\begin{abstract}
We use a magnetic force microscope (MFM) to investigate single vortex pinning and penetration depth in \Nd, one of the highest-$T_c$ iron-based superconductors. In fields up to 20 Gauss, we observe a disordered vortex arrangement, implying that the pinning forces are stronger than the vortex-vortex interactions. We measure the typical force to depin a single vortex, $F_{\mathrm{depin}} \simeq 4.5$ pN, corresponding to a critical current up to $J_c \simeq 7\times 10^5$ A/cm$^2$. Furthermore, our MFM measurements allow the first local and absolute determination of the superconducting in-plane penetration depth in \Nd, $\lambda_{ab}=320\pm60$ nm, which is larger than previous bulk measurements.
\end{abstract}

\maketitle

\section{\label{sec:intro}Introduction}

A long-standing challenge on the road to superconducting applications is the unwanted motion of vortices, quanta of magnetic flux $\Phi_0=h/2e$ which penetrate a type-II superconductor in the presence of a magnetic field. Each vortex consists of a circulating supercurrent which decays radially on the length scale of the penetration depth $\lambda$, and a core of suppressed superconductivity with radius given by the coherence length $\xi_0$. When a current is applied to the superconductor, the vortices experience a Lorentz force density $\vec{F}_L = \vec{J} \times \vec{B}$, perpendicular to the direction of the current and proportional to its magnitude. Although the net supercurrent remains non-dissipative, the accompanying motion of the normal electrons in the vortex cores causes resistance and energy loss in the material.

A vortex can be pinned by co-locating its energetically costly core with a preexisting crystal defect where superconductivity is already suppressed. The strength of the pinning force density $F_P$ dictates the maximum supercurrent which can be applied without vortex motion and consequent dissipation. Several decades of engineering effort have been devoted to optimizing vortex pinning in superconductors.\cite{LarbalestierNature2001, ScanlanIEEE2004, FoltynNatMat2007} However, vortex pinning in the highest-$T_c$ cuprate superconductors remains challenging, due in part to the large electronic anisotropy which allows vortices to bend and depin in pancake fragments rather than as one-dimensional semi-rigid objects.\cite{ClemPRB1991}


The recent discovery of high-$T_c$ iron-based superconductors (Fe-SCs) brought new optimism to the vortex pinning problem.\cite{PuttiSST2010, ZhangFrontPhys2012} Fe-SCs were found to be typically more isotropic than their cuprate cousins.\cite{JiaAPL2008, YamamotoAPL2009} Furthermore, intrinsic pinning tests showed promise,\cite{GaoSST2008} raising hopes that defect engineering could improve the pinning properties to achieve critical currents larger than the cuprate benchmark around $10^6$\,A/cm$^2$.\cite{FoltynNatMat2007}

Maximizing critical current will require detailed understanding of the pinning efficacy of specific defects. Transport measurements yield information about macroscopic critical currents but cannot directly distinguish the distribution of pinning forces and the sites responsible for the strongest pinning. Furthermore, the dissipative motion of small numbers of vortices may not be detected by macroscopic transport measurements. It is crucial to understand and quantify pinning at the single vortex level. An important parameter in this endeavor is $\lambda$, the fundamental length scale of magnetic interactions, whose absolute value has remained challenging to measure via bulk techniques due to pervasive inhomogeneity in Fe-SCs.



While scanning tunneling microscopy,\cite{YinPRL2009, YinPhysicaC2009, HanaguriScience2010, ShanNatPhys2011, SongScience2011, HanaguriPRB2012, WangNatPhys2012, SongPRL2012, SongPRB2013} Bitter decoration,\cite{EskildsenPRB2009, EskildsenPhysicaC2009, VinnikovJETP2009, LiPRB2011, DemirdisPRB2011} and scanning SQUID\cite{HicksJPSJ2008, KaliskyPRB2010, KaliskyPRB2011} techniques can image vortex configurations and offer some insight into pinning, magnetic force microscopy (MFM) is the most powerful technique to directly quantify the single-vortex depinning force $F_{\mathrm{depin}}$ and the in-plane penetration depth $\lambda_{ab}$. MFM was used to measure vortex lattice correlation length, $F_{\mathrm{depin}}$, and $\lambda_{ab}$ distributions in doped BaFe$_2$As$_2$,\cite{LuanPRB2010, LuanPRL2011, InosovPRB2010, YangPRB2011, KimSST2012, LamhotPRB2015} but these parameters have not been locally quantified in the highest-$T_c$ `1111' family of Fe-SCs. More details of previous work are summarized in Appendix \ref{app:PreviousResults}. Here we use MFM to investigate \Nd. We find a typical single-vortex depinning force $F_{\mathrm{depin}} \simeq 4.5$\,pN and penetration depth $\lambda_{ab}=320\pm60$\,nm. The tip-induced vortex motion demonstrates out-of-plane electronic anisotropy, but no statistically significant evidence of in-plane anisotropy in \Nd.

\section{\label{sec:experiment}Experiment}

\begin{figure}
\center
\includegraphics[width=1\columnwidth]{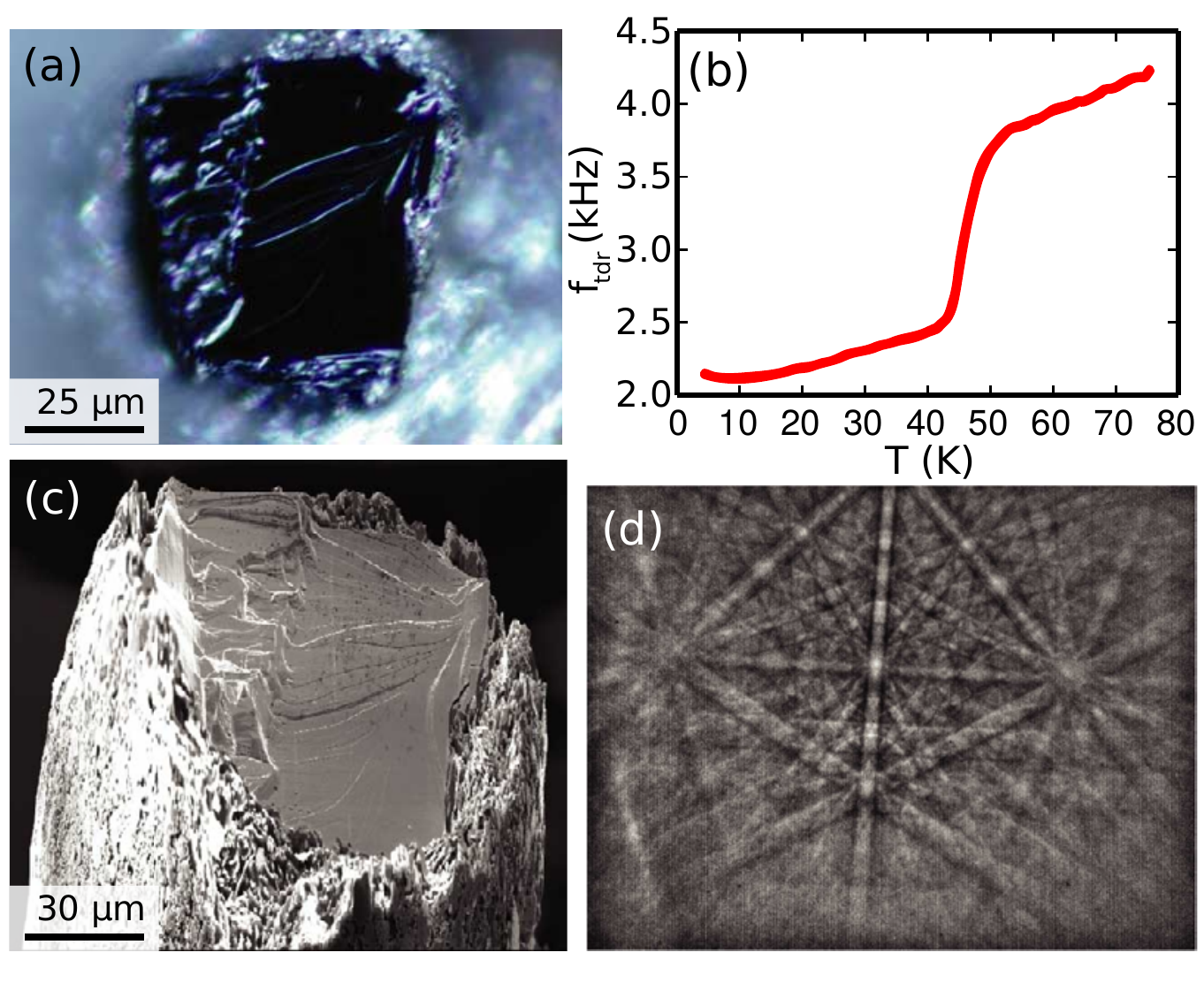}
\caption{Sample characterization. (a) Photograph of the \Nd\ single crystal. (b) Tunnel diode resonator experiment shows $T_c(\mathrm{onset})\sim 50$\,K. (c) SEM image of the same crystal. (d) Room temperature EBSD Kikuchi patterns determining the axis orientation. This sample is expected to remain in the tetragonal structure for all temperatures,\cite{QiuPRL2008} but our EBSD data lacks the resolution to independently distinguish between tetragonal ($P4/nmm$) and orthorhombic ($Cmma$) structures.}
\label{fig:OrientationTc}
\end{figure}

The single crystal used in this experiment, of thickness $\sim$10 $\mu$m and lateral dimension $\sim$100 $\mu$m, shown in Fig.\ \ref{fig:OrientationTc}(a), was mechanically extracted from polycrystalline \Nd\ synthesized at high pressure.\cite{KondoPRL2008} It is slightly underdoped with nominal $x=0.1$, and $T_c(\mathrm{onset}) \simeq 50$\,K measured by tunnel diode resonator (Fig.\ \ref{fig:OrientationTc}(b)). The sample was cleaved in air perpendicular to the $c$-axis, aligned in the MFM, pumped to high vacuum, cooled to base temperature $\sim$6 K, and imaged in cryogenic ultra-high vacuum (UHV) at applied fields up to $\pm20$ Gauss. After the MFM experiment, the sample crystallinity and orientation were characterized by scanning electron microscopy (SEM) and electron backscatter diffraction (EBSD) as shown in Figs.\ \ref{fig:OrientationTc}(c-d).

MFM imaging was carried out in frequency modulation mode\cite{AlbrechtJAP1991} using a commercial NSC18 cantilever from Mikromasch with manufacturer-specified force constant $k=3.5\pm 1.5$ N/m.\cite{cantilever} As the tip oscillates in the $z$ axis, the magnetic moment at the end of the tip interacts with the vortex. The bare cantilever resonance frequency $f_0 = 73$ kHz is shifted by the local force gradient according to
\begin{equation}
df \approx -\frac{f_0}{2k} \frac{\partial F_z(x,y,z)}{\partial z} .
\label{eq:freqshift}
\end{equation}
\noindent The total force consists primarily of van der Waals and magnetic components, with the former dominant when the tip is close to the sample.\cite{HaberleNJP2012} We use the rapid change in $dF_z/dz$ to determine the position of the sample surface to within a few nanometers, then retract the tip by a fixed amount $z$ for constant height imaging in a regime where the magnetic force is dominant.

We employ a feedback loop to maintain the tip oscillation at its local resonance frequency as the sample is scanned; the resultant map of the local frequency shift $df$ thus serves as a map of the vertical force gradient, $\partial F_z/ \partial z$. $F_z$ is not directly imaged, but it can be obtained by integrating the measured $\partial F_z/ \partial z$. The lateral force components can then be estimated by assuming a truncated cone tip shape.\cite{LuanPRB2010} Although Newton's third law, which dictates that any force used as an imaging signal also influences the sample itself, is typically regarded as a disadvantage which makes a force microscope an invasive probe, we use it here to our advantage in order to pull or push on vortices and directly measure the forces required to dislodge them from their pinning sites in \Nd. More details of the tip-vortex interaction regimes are given in Appendix \ref{app:VortexRegimes}.

\begin{figure}[t]
\center
\includegraphics[width=1\columnwidth]{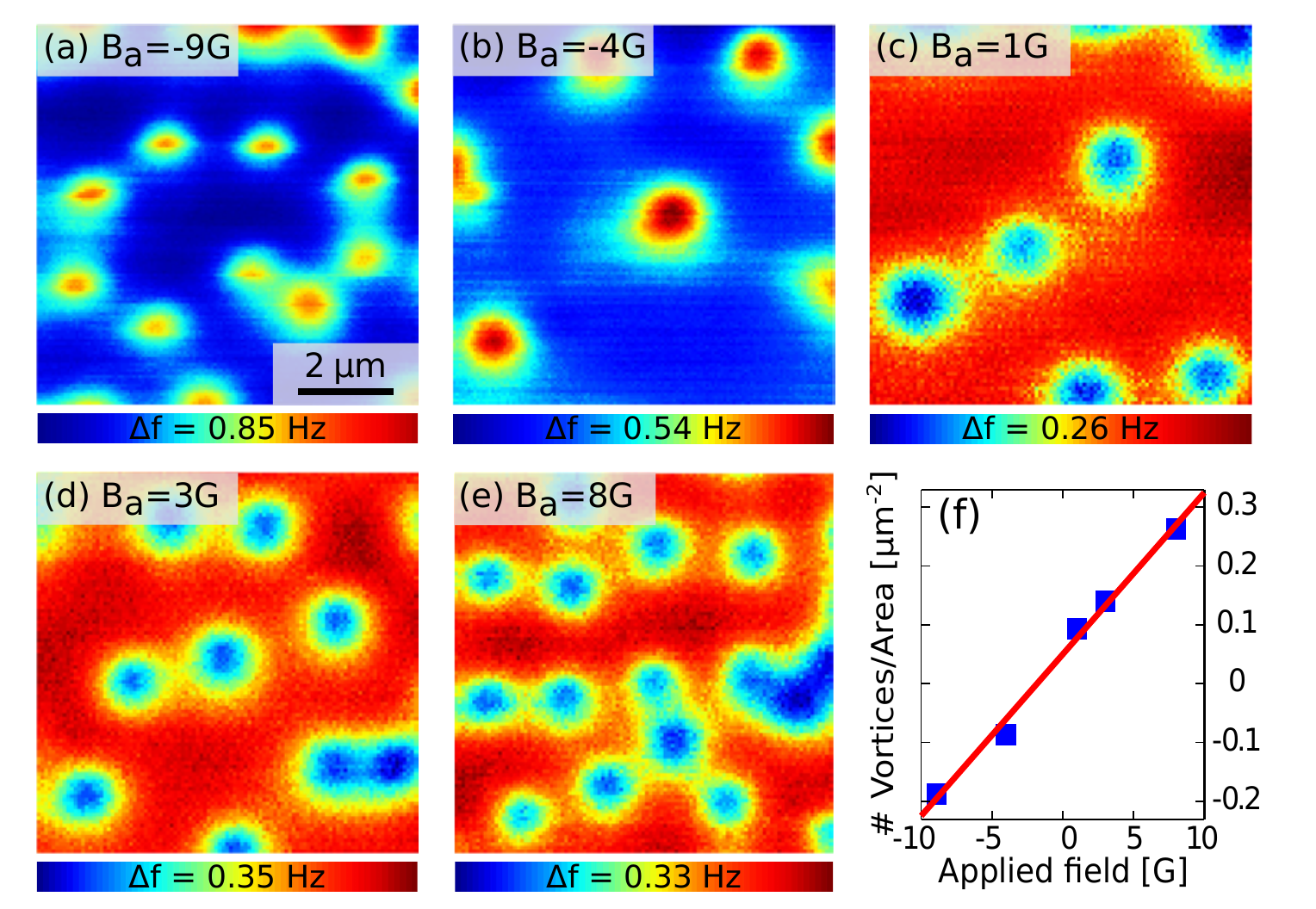}
\caption{MFM images of vortices vs.\ applied field $B_a$ in \Nd. (a) $B_a=-9$ G ($z=200$ nm). (b) $B_a=-4$ G ($z=300$ nm). (c) $B_a=1$ G ($z=400$ nm). (d) $B_a=3$ G ($z=300$ nm). (e) $B_a=8$ G ($z=300$ nm). (f) Vortex density vs.\ $B_a$. At these scan heights, the tip-vortex force does not cause significant vortex depinning.}
\label{fig:BvsN}
\end{figure}

\section{\label{sec:results}Results}

\subsection{\label{sec:image}Vortex Imaging}

\begin{figure*}[ht]
\center
\includegraphics[width=1.6\columnwidth]{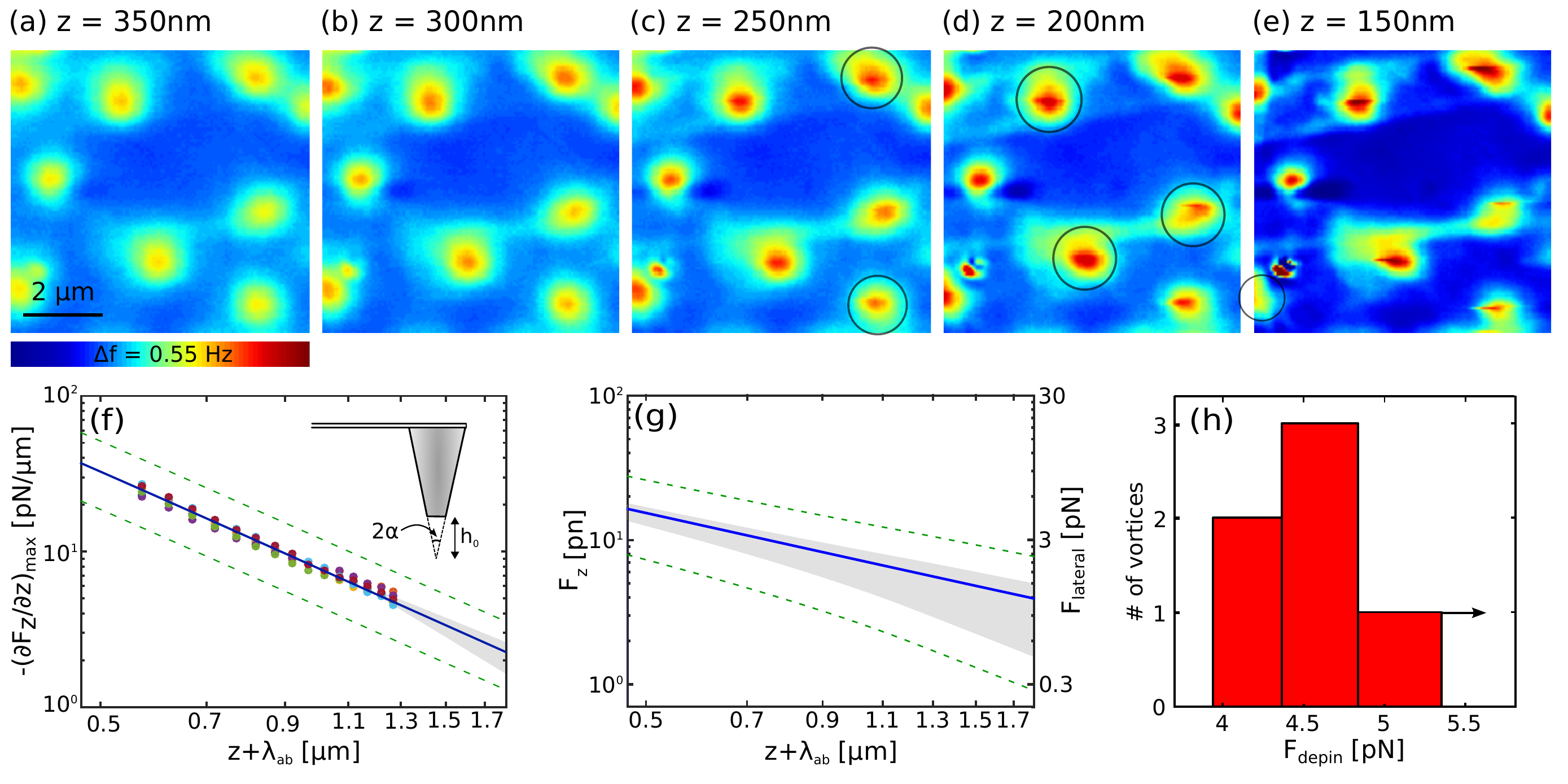}
\caption{Vortex depinning. (a-e) MFM images of the same field of view containing $\sim$9 vortices at $T=6$ K and $B=5.8$ Gauss, with decreasing tip-sample separation $z$. As $z$ decreases, the increasing lateral force between tip and sample causes some vortices to depin. Black circles indicate vortices depinning for the first time. (f) Maximum vertical force gradient $(dF_z/dz)_{\mathrm{max}}$ vs.\ $z+\lambda_{ab}$ (with $\lambda_{ab}=320$ nm) for the 7 vortices entirely within the field of view. The values for each vortex are depicted as a different series of colored dots, terminating at the minimum $z$ where the vortex remains pinned. Blue line shows a fit to Eq.\ \ref{eq:tipVortex} with fixed $h_0=200$\,nm and free parameters $A$ and $\lambda_{ab}$. The gray shading denotes the region between $(z+\lambda_{ab})^{-3}$ and $(z+\lambda_{ab})^{-2}$, which is used to bound any systematic error in the truncated cone tip model. Green dashed lines indicate the other major source of systematic error, the uncertainty in the manufacturer-specified spring constant $k$. Inset shows a sketch of the magnetic tip, which defines $h_0$. (g) The integral of (f), using the functional form of the fit to extrapolate to $z=\infty$, gives the total $F_z^{\mathrm{max}}$, shown as a blue line. Gray shading and green dashed lines define the systematic errors from the model and the spring constant, respectively. (h) Histogram of $F_{\mathrm{depin}}$ for the 6 displaced vortices shown in (a-e).}
\label{fig:PinningForce}
\end{figure*}

Figs.\ \ref{fig:BvsN}(a-e) show images of individual vortices vs.\ applied field $B_a$ in \Nd. Before the acquisition of each image, the sample was heated above $T_c$, to 60 K, then cooled to 6 K within the field shown. We note immediately that the vortices are disordered, indicating that at this low density the pinning forces on individual vortices dominate over the vortex-vortex interaction forces in this \Nd\ sample. The vortex density vs.\ applied field is shown in Fig.\ \ref{fig:BvsN}(f). The intercept serves to calibrate the ambient $B$ field offset, while the slope serves to calibrate the $(x,y)$ scan piezo. We find an offset field of 1.8 Gauss, consistent with Earth's field and stray fields from steel screws close to the MFM.
For all following figures and discussion, we employ the calibrated $B$ and $(x,y)$ distances.


\subsection{\label{sec:pin}Pinning Force Distribution}

We measure the vortex pinning force distribution by acquiring a series of images starting at tip height $z_{\mathrm{max}}=950$\,nm, and approaching the sample, as exemplified in Figs.\,\ref{fig:PinningForce}(a-e).  As the tip-vortex force increases in successive scans, we note the height $z_{\mathrm{depin}}$ of the first depinning event for each vortex (marked with a circle). To obtain the force corresponding to $z_{\mathrm{depin}}$, we first plot in Fig.\ \ref{fig:PinningForce}(f) the maximum force gradient $(dF_z/dz)_{\mathrm{max}}$ at each vortex center. To compute $F_z$ at these locations, we must integrate $dF_z/dz$ from $z=\infty$ to $z_{\mathrm{depin}}$. Between $z_{\mathrm{max}}$ and $z_{\mathrm{pin}}$, we simply sum the measured $(dF_z/dz)_{\mathrm{max}}$. To handle the tail from $z=\infty$ to $z_{\mathrm{max}}$, we model the vortex as a monopole situated $\lambda_{ab}$ beneath the surface.\cite{PearlJAP1966} We model the tip as a uniformly coated truncated cone, depicted as an inset to Fig.\ \ref{fig:PinningForce}(f),\cite{LuanPRB2010} which leads to
\begin{equation}
\left( \frac{dF_z}{dz} \right)_{\mathrm{max}}= \Phi_0 A \left[ \frac{1}{(z+\lambda_{ab})^2} + \frac{2h_0}{(z+\lambda_{ab})^3} \right].
\label{eq:tipVortex}
\end{equation}
\noindent Here $h_0$ is the tip truncation height (which we fix at 200 nm based on manufacturer specifications) and $A=\alpha t m$ is a tip-specific fit parameter depending on the opening angle $\alpha$, the magnetic film thickness $t$, and the magnetization $m$. We fit the measured $(dF_z/dz)_{\mathrm{max}}$ vs.\ $z$, extrapolate the fit function to $z=\infty$, integrate the tail, and add this result to the sum of the measured $(dF_z/dz)_{\mathrm{max}}$, to arrive at the total $F_z^{\mathrm{max}}$, shown in Fig.\ \ref{fig:PinningForce}(g). We take $F_{\mathrm{depin}}$ to be the maximum lateral force $F_{xy}^{\mathrm{max}}$, which is a factor of $\sim0.31$ times $F_z^{\mathrm{max}}$ for a truncated cone tip. Finally, Fig.\ \ref{fig:PinningForce}(h) shows a histogram of $F_{\mathrm{depin}}$ for the six vortex motion events within this field of view. Details of the analysis and error bars are given in Appendix \ref{app:PinningForceDetails}.

From Fig.\ \ref{fig:PinningForce}(h), the typical single-vortex depinning force is $F_{\mathrm{depin}} \simeq 4.5$ pN. In order to convert the pinning force to a critical current $J_c = F_{\mathrm{depin}}/(\Phi_0 \ell)$, we need an estimate of the length $\ell$ of the vortex which is depinned. Although the sample thickness is on order 10 $\mu$m, which places an upper bound on $\ell$, our observation of vortex wiggling in Fig.\ \ref{fig:InteractionRegimes}(b), as well as the occasional observation of a flux `tail' after a vortex dragging event in Fig.\ \ref{fig:InteractionRegimes}(i), suggests the possibility that only the top portion of the vortex is depinned by the tip. The visibility of the stationary bottom portion of the vortex decays exponentially with its distance beneath the surface,\cite{GuikemaPRB2008, LuanPRB2009} placing an approximate lower bound of $\lambda$ on the length of the upper displaced portion. The measured $F_{\mathrm{depin}} \simeq 4.5$ pN thus corresponds to $4.5\times 10^{-7}$ to $1.4 \times 10^{-5}$ N/m.

The measured depinning force is roughly consistent with the density $\sim$2-60$\times 10^{-22}$ m$^{-3}$ of strong pinners with individual strength $2\times 10^{-13}$ N per pinner previously estimated by magneto-optical imaging on underdoped \Nd\ ($T_c \sim 35$\,K).\cite{VanderBeekPRB2010} Our measurement therefore implies a critical current $2\times 10^4$ A/cm$^2 \lesssim J_c \lesssim 7\times 10^5$ A/cm$^2$. These values greatly exceed the bulk critical current density $J_c\sim 2000$ A/cm$^2$ in polycrystalline \Nd,\cite{YamamotoSST2008} and are roughly consistent with the in-grain persistent current density detected by magneto-optical imaging, $J_c \sim 10^5$ A/cm$^2$ (Refs.\ \onlinecite{ProzorovNJP2009} and \onlinecite{VanderBeekPRB2010}) and $J_c \sim 5\times 10^6$ A/cm$^2$ (Ref.\ \onlinecite{YamamotoSST2008}), and with requirements for various superconducting applications.\cite{LarbalestierNature2001}

\subsection{\label{sec:lambda}Penetration Depth}

The superconducting penetration depth $\lambda$ is necessary to quantitatively model the single vortex pinning force, as well as to determine fundamental parameters such as the phase stiffness and superfluid density. The temperature dependence of $\lambda(T)$ can be used to distinguish between fully-gapped vs.\ nodal pairing scenarios.\cite{MartinPRL2009, LuanPRB2010}
However, the absolute value of $\lambda$ has been notoriously difficult to measure in 1111 Fe-SCs via bulk techniques, due to prevalent sample inhomogeneity.\cite{ProzorovROPP2011}

Two common techniques to measure $\lambda$ are muon spin rotation ($\mu$SR) and lower critical field ($H_{c1}$) measurements. First, $\mu$SR experiments measure a histogram of the spin decay times of stopped muons within a superconducting sample, which translates to a histogram of the local magnetic field strengths present in the sample. Assuming a triangular vortex lattice in NdFeAsO$_{0.85}$ with $T_c=51$ K, one $\mu$SR study extracted $\lambda_{ab}=195\pm 5$ nm,\cite{KhasanovPRB2008} but our observation of a disordered vortex arrangement in \Nd\ suggests that this $\mu$SR study underestimates $\lambda_{ab}$. Second, $\lambda_{ab}$ can be related to $H^c_{c1}$ via $H^c_{c1}=\Phi_0/(4\pi\lambda_{ab}^2)\left[\ln\kappa+c(\kappa)\right]$, where $\kappa=\lambda_{ab}/\xi_{ab}$ and $c(\kappa)$ is a function that tends to 0.5 for large values of $\kappa$. A measurement of $H^c_{c1}=150$ Gauss and $\xi_0\sim2.9$ nm (from $H_{c2}=40$ T) can be used to determine $\lambda_{ab}=270\pm40$ nm in \Nd\ with $T_c=32.5$ K.\cite{PribulovaPRB2009}

We use MFM to measure $\lambda_{ab}$ locally, via the Meissner repulsion of the tip in a vortex-free region. For $z \gtrsim \lambda_{ab}$, the tip-superconductor interaction can be approximated by the interaction between the tip and an image tip reflected across a mirror plane a distance $\lambda_{ab}$ below the surface of the superconductor.\cite{LuanPRB2010} For $z > 300$ nm, we fit to the Meissner repulsion
\begin{multline}
\frac{\partial F_z}{\partial z} - \left. \frac{\partial F_z}{\partial z} \right|_{z=\infty} \\
= 2\pi \mu_0 A^2 \left[ \frac{1}{z+\lambda_{ab}} + \frac{h_0}{(z+\lambda_{ab})^2} + \frac{h_0^2}{2(z+\lambda_{ab})^3} \right]
\label{eq:tipMeissner}
\end{multline}
\noindent to find $\lambda_{ab}=320\pm60$\,nm. Details of the analysis are given in Appendix \ref{app:PenetrationDepthDetails}.

From our absolute measurement of $\lambda_{ab}$, we can compute several fundamental quantities. Using $k_B T_F = \hbar^2 \pi c_{\mathrm{int}} n_s/m^{*}$ and $1/\lambda_{ab}^2 = 4\pi e^2/c^2 \cdot n_s/m^{*}$,\cite{UemuraPRL1991} where $c_{\mathrm{int}}=8.557$ \AA\ is the distance between
superconducting layers,\cite{QiuPRL2008} we compute $T_F=650$ K, which is in rough agreement with the trend $T_F \sim 10 T_c$ noted for a variety of unconventional superconductors.\cite{UemuraPRL1991}

\begin{figure}
\center
\includegraphics[width=1\columnwidth]{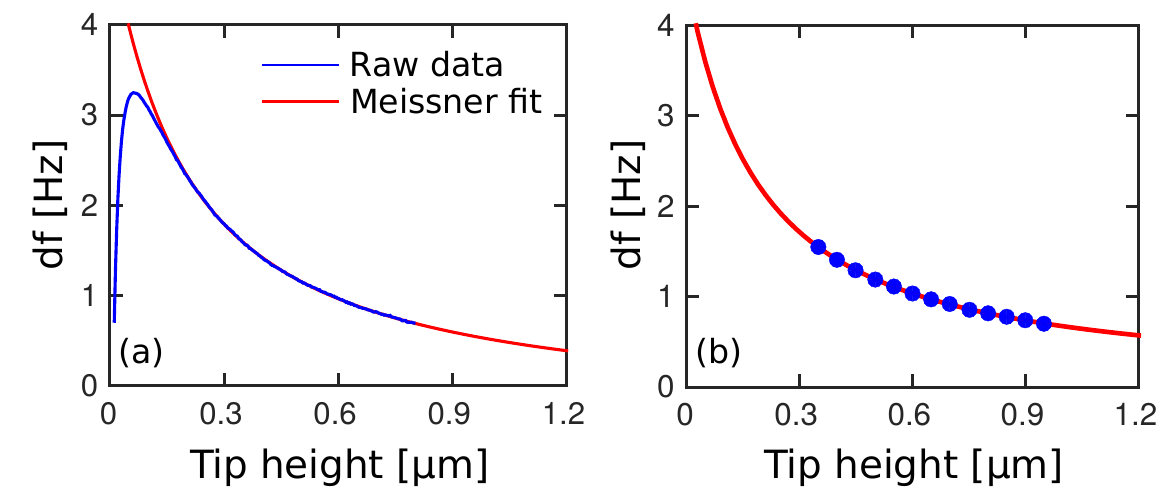}
\caption{Penetration depth. (a) The blue line shows $d f$ vs.\ tip-sample distance $z$ at a location which is laterally at least 2 $\mu$m from the nearest vortex ($T = 6$ K; $B = 1$ Gauss.) The red line shows a fit to Eq.\,\ref{eq:tipMeissner} for $z > 300$ nm. (b) Blue dots show the background values of $d f$ from the vortex images in Fig.\ \ref{fig:PinningForce}. The red line is a fit to Eq.\,\ref{eq:tipMeissner} for $z > 300$ nm.
\label{fig:MeissnerLambda}}
\end{figure}

\section{\label{sec:conclusion}Conclusion}

In summary, we have presented  a systematic investigation of single vortex pinning in one of the highest-$T_c$ Fe-SCs, \Nd, for $B$ up to 20 Gauss.  We find a disordered vortex arrangement, implying that the pinning forces are stronger than the vortex-vortex interactions. The average pinning force of a single vortex is $F_P \simeq 4.5$ pN, consistent with a critical current $j_c$ up to $7\times 10^5$ A/cm$^2$, within striking distance of the benchmark value for cuprates. \Nd\ presents a rich opportunity for vortex physics, with the out-of-plane anisotropy sitting right at the boundary between pancake and rigid Abrikosov vortices. (No in-plane anisotropy was detected, as detailed in Appendix \ref{app:Anisotropy}.) \Nd\ also presents significant technical opportunity to improve the inter-grain $j_c$ by the reduction of wetting impurity phases\cite{KametaniSST2009} and the intra-grain $j_c$ by the improvement of nanoscale pinning defects. The ability to intentionally position the vortex top using the MFM tip opens the door for experiments to measure pinning forces at specific sample defects which may be independently imageable in AFM mode (i.e.\ with smaller tip-sample separation $z$, where topographic and electrostatic forces dominate over magnetic ones).

\begin{acknowledgments}
 We thank Matthew Tillman for assistance with the crystal growth, Hyun Soo Kim for the tunnel diode resonator characterization in Fig.\ \ref{fig:OrientationTc}(b), and Natasha Erdman for the EBSD measurements in Figs.\ \ref{fig:OrientationTc}(c) and (d). We thank Ilya Vekhter and Peter Hirschfeld for helpful conversations about vortex pinning anisotropy. Crystal growth and characterization at Ames Laboratory were carried out by R.P., S.K., and P.C.C., with support from the U.S. Department of Energy, Office of Basic Energy Science, Division of Materials Sciences and Engineering. Ames Laboratory is operated for the U.S. Department of Energy by Iowa State University under Contract No. DE-AC02-07CH11358. The MFM experiment was carried out by J.K. and supported by Harvard's Nanoscale Science and Engineering Center, funded by NSF grant PHY 01-17795. The collaborative data analysis was carried out by J.Z., M.H., C.Y., O.M.A., and J.E.H., with support from the U.S.-Israel Binational Science Foundation under grant number 2010305. M.H.\ also acknowledges the support of the Deutsche Forschungsgemeinschaft (HU 1960/11) and J.Z.\ acknowledges support from MIT's UROP program.
\end{acknowledgments}

\appendix

\section{\label{app:PreviousResults}Previous Vortex Imaging Studies}

Single vortices have been directly imaged in all four major families of Fe-SCs. Due to the relative ease of large single crystal growth, the `122' family has been the most heavily investigated. Scanning tunneling microscopy (STM) first showed a disordered vortex lattice indicative of strong pinning in optimally doped Ba(Fe$_{0.9}$Co$_{0.1}$)$_2$As$_2$ with $T_c = 25.3$ K.\cite{YinPRL2009} Furthermore, the lack of correlation between vortex locations and surface defects suggested that vortices behave as semi-rigid one-dimensional objects,\cite{YinPhysicaC2009} consistent with low anisotropy $\gamma=H_{c2}^{ab}/H_{c2}^c \lesssim 2$ measured by transport techniques.\cite{YamamotoAPL2009} Bitter decoration imaging of larger arrays of vortices in the same material confirmed the vortex disorder,\cite{EskildsenPRB2009, EskildsenPhysicaC2009, VinnikovJETP2009} while magnetic force microscopy (MFM) quantified the scaling of vortex disorder with applied field.\cite{InosovPRB2010} In underdoped Ba(Fe$_{0.95}$Co$_{0.05}$)$_2$As$_2$ ($T_c=18.5$ K), MFM imaging demonstrated the uniformity of the superfluid density at the length scale of the penetration depth $\lambda_{ab}=325\pm50$ nm, and allowed measurement of a typical single-vortex depinning force $F_{\mathrm{depin}}=18$ pN at $T=5$ K in a twin-free, 10 $\mu$m-thick sample, corresponding to $1.8\times 10^{-6}$ N/m.\cite{LuanPRB2010} Another Bitter decoration study of under- and optimally-doped Ba(Fe$_{1-x}$Co$_{x}$)$_2$As$_2$ found an even larger typical pinning force of order $10^{-5}$ N/m.\cite{DemirdisPRB2011} In contrast, STM imaging of optimally doped K$_{0.4}$Ba$_{0.6}$Fe$_2$As$_2$ with $T_c=38$ K showed an ordered vortex lattice indicative of weak pinning.\cite{ShanNatPhys2011} MFM imaging of optimal and underdoped K$_{x}$Ba$_{1-x}$Fe$_2$As$_2$ showed disorder at low applied fields, with a tendency towards an ordered lattice at Tesla fields.\cite{YangPRB2011} Furthermore, the low-field (10 Gauss) pinning force was measured to be $2\times 10^{-7}$ N/m,\cite{YangPRB2011} a full order of magnitude weaker than in Ba(Fe$_{0.95}$Co$_{0.05}$)$_2$As$_2$.\cite{LuanPRB2010} A possible resolution to the pinning differences within the `122' family was prompted by the observation of disordered vortices K$_x$Sr$_{1-x}$Fe$_2$As$_2$:\cite{VinnikovJETP2009, SongPRB2013} dopant size mismatch and consequent clustering may increase electronic inhomogeneity at the $\xi_0$ length scale and enhance vortex pinning.\cite{SongPRB2013} Very disordered vortices were also observed by Bitter decoration in Ba(Fe$_{1-x}$Ni$_x$)$_2$As$_2$, with a tendency to form vortex stripes in some regions.\cite{LiPRB2011} Finally, MFM imaging of vortices in Ba(As$_{1-x}$P$_{x}$)$_{2}$Fe$_{2}$ showed relatively strong, uniform pinning for overdoped samples, clustered pinning along twin boundaries for slightly underdoped samples, and weak pinning for very underdoped samples.\cite{LamhotPRB2015}

In the `11' family, STM imaging showed a disordered vortex arrangement in FeSe$_{x}$Te$_{1-x}$ with $x\sim0.4$,\cite{HanaguriScience2010} but an ordered vortex lattice in stoichiometric FeSe films,\cite{SongScience2011} consistent with the idea that dopant clustering may cause the inhomogeneity necessary for pinning.\cite{SongPRB2013}

In the `111' family, STM imaging of LiFeAs found vortex disorder increasing with applied field,\cite{HanaguriPRB2012} which can be attributed to a decreasing shear modulus $C_{66}$ that allows pinning forces to dominate over the vortex-vortex interaction forces.\cite{BrandtPRB1986} STM imaging of vortices in Na(Fe$_{0.975}$Co$_{0.025}$)As at 11 T also showed a slightly disordered lattice.\cite{WangNatPhys2012}

Scanning squid microscopy images of vortices in twinned Ba(Fe$_{1-x}$Co$_x$)$_2$As$_2$ showed that vortices avoid twin boundaries,\cite{KaliskyPRB2011} consistent with a report that twin boundaries enhance the local superfluid density.\cite{KaliskyPRB2010} In contrast, twin boundaries were found to pin vortices in FeSe.\cite{SongPRL2012} Taken together, all of these studies show diverse pinning behavior across the `122', `11', and `111' families of Fe-SCs.

In the `1111' family, scanning SQUID microscopy showed disordered vortices in polycrystalline \Nd, but the vortex profiles were severely resolution-limited.\cite{HicksJPSJ2008} Bitter decoration of single crystalline SmFeAsO$_{1-x}$F$_x$ at 46 Gauss also showed disordered vortices, with some tendency to form stripes, but the images were again severely resolution-limited.\cite{VinnikovJETP2009} Neither the intrinsic vortex profile, nor the single vortex pinning forces have ever been quantified in this highest-$T_c$ family of Fe-SCs.

\section{\label{app:VortexRegimes}Tip-Vortex Interaction Regimes}

\begin{figure*}
\center
\includegraphics[width=1.5\columnwidth]{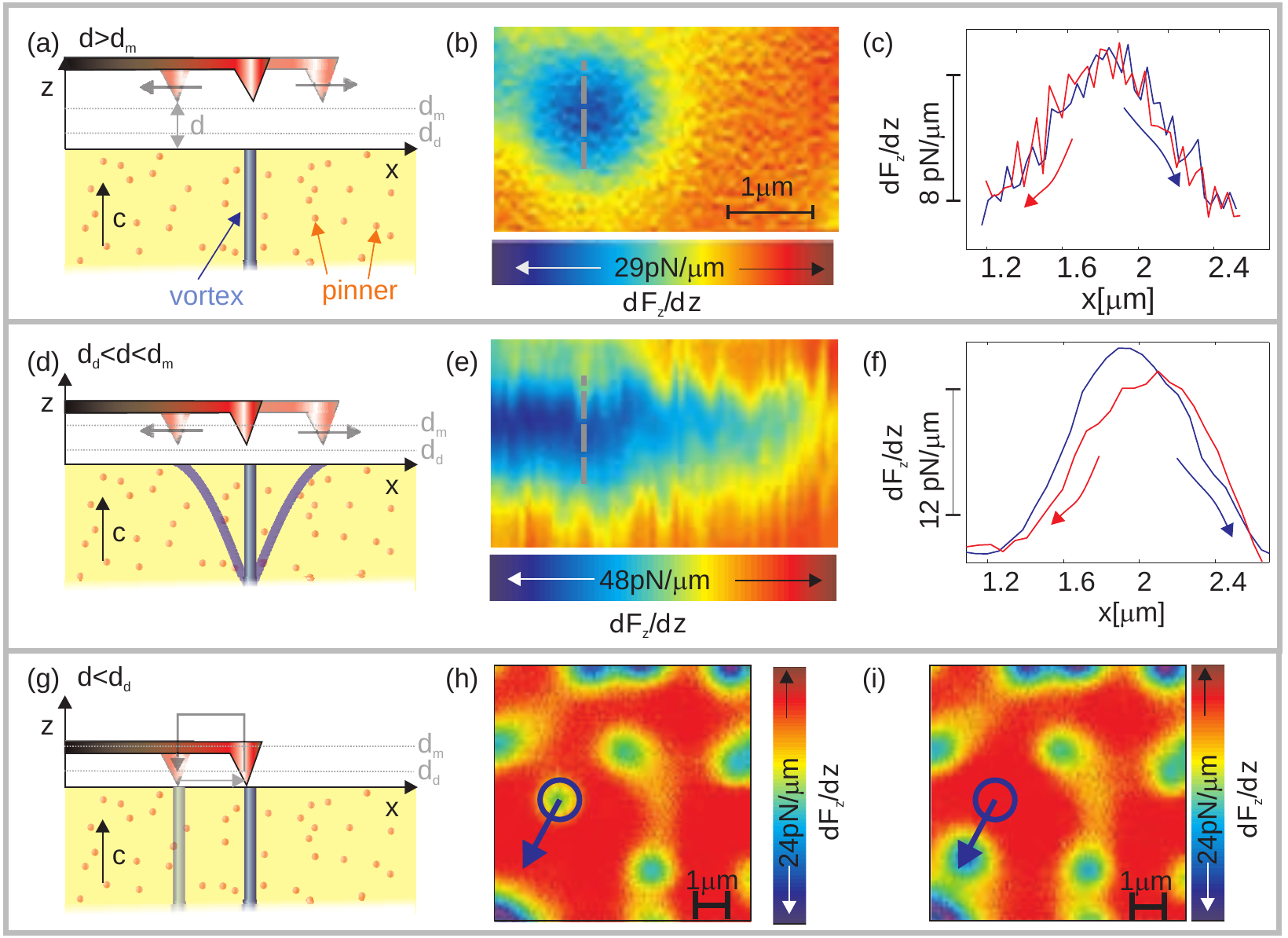}
\caption{(a-c) Large $z$: passive observation. (d-f) Intermediate $z$: only the top of each vortex is displaced, as it stretches away from an anchor point deeper within the crystal. (g-i) Small $z$: in some cases, the entire observable vortex is displaced. Note that one of the vortices has a `tail' indicating that only the top portion of length $\sim\lambda$ was displaced with respect to the bottom portion.\cite{GuikemaPRB2008, LuanPRB2009}}
\label{fig:InteractionRegimes}
\end{figure*}

We tune the tip-vortex force by varying the tip height $z$, as shown in Fig.\ \ref{fig:InteractionRegimes}. At large $z$ (`surveillance height'), the interaction force is small, so the MFM can be used to image vortices without displacing them, as shown in Figs.\ \ref{fig:InteractionRegimes}(a-c).

As we lower the tip height to a regime which exceeds most pinning forces, we observe that most vortices wiggle back and forth as the tip is rastered over them, as shown in Figs.\ \ref{fig:InteractionRegimes}(d-f). This behavior implies that the vortex is bending at the top, as sketched in Fig.\ \ref{fig:InteractionRegimes}(d).\cite{AuslaenderNatPhys2008}

As we lower the tip still further, turning off the oscillation and approaching to within 50 nm of the sample, the tip-sample force becomes strong enough to interact with a deeper portion of the vortex, and the full observable top of the vortex can be moved as a single unit, as shown in Fig.\ \ref{fig:InteractionRegimes}(g-i). However, we do occasionally see weak tails which indicate that even in this interaction mode, the top portion of the vortex may decouple from the lower portion, suggesting significant out-of-plane anisotropy.

The energy cost of an Abrikosov vortex scales with the volume of its core and the free energy of the superconductor. To minimize the core volume, the vortex will take the shortest path through the sample, deviating only slightly to take advantage of pinning sites along the way. However in an anisotropic layered material a vortex may kink its way along an indirect path through the sample like an offset stack of pancakes.\cite{ClemPRB1991} Each pancake can pin independently, and the tip interacts with each pancake with exponentially decaying force.\cite{GuikemaPRB2008, LuanPRB2009} In \Nd\ with nominal $x=0.18$ and $T_c(\mathrm{onset})=50$ K, transport measurements on single crystals in field yield anisotropy $\gamma=H_{c2}^{ab}/H_{c2}^{c} \leq 4.34-4.9$ while cleaner crystals with $T_c(\mathrm{onset})=52$ K yield $\gamma \leq 6$ (Ref.\ \onlinecite{JiaAPL2008}). On another sample of \Nd\ with $T_c\sim34$ K, the anisotropy was $\gamma\sim 5.5$ from the ratio of the upper critical fields, $\gamma\sim 4$ from the ratio of the lower critical fields,\cite{PribulovaPRB2009} and $\gamma \sim 7.5$ from the ratio of the irreversibility line.\cite{KacmarcikPRB2009}

\section{\label{app:PinningForceDetails}Pinning Force Analysis}

The magnetic field of a vortex at a distance $z$ above the $ab$ surface of a superconducting half-space is\cite{KoganPRB1993, LuanThesis2011}
\begin{equation}\label{eq_total_B}
\vec{B}(\vec{R},z)=\frac{\Phi_0}{(2\pi)^2}\int d^2q\frac{e^{i\vec{q}\cdot\vec{R}-qz}}{Q(Q+\lambda_{ab} q)}(\hat{z}-i\hat{q})\end{equation}
where $\lambda_{ab}$ is the in-plane penetration depth and $Q=\sqrt{1+(\lambda_{ab} q)^2}$.
The total interaction energy between an MFM tip and a vortex is
\begin{equation}\label{eq:tip_vortex_energy}
U(\vec{R},z) = \int_{\mathrm{tip}}dz'd^2R' M_z(\vec{R'},z') B_z(\vec{R}+\vec{R'},z+z')
\end{equation}
where $M_z(\vec{R'},z')$ is the magnetic density distribution, and the primed coordinates refer to the tip, with origin at the center of the plane which truncates the conical tip. The signal measured by the MFM is the vertical gradient of the magnetic force, which is given by
\begin{equation}\label{eq:vor_lateral_exact}
\frac{\partial F_z}{\partial z}(R,z)=\int_{\mathrm{tip}} dz'd^2R' M_z(\vec{R'},z')\frac{\partial^2}{\partial z'^2}B_z(\vec{R}+\vec{R'},z+z').
\end{equation}

For the truncated cone tip model with small opening angle $\alpha$, Eq.\ \ref{eq:vor_lateral_exact} can be approximated by
\begin{multline}\label{eq:vor_lateral}
\frac{\partial F_z}{\partial z}(R, z) \\
 \approx \Phi_0 A\left[\frac{z+\lambda_{ab}}{((z+\lambda_{ab})^2+R^2)^{3/2}} + \frac{h_0(2(z+\lambda_{ab})^2-R^2)}{((z+\lambda_{ab})^2+R^2)^{5/2}}\right]
\end{multline}
\noindent where $A=\alpha t m$. Setting $R=0$ in equation (\ref{eq:vor_lateral}) gives the peak vertical force gradient above a vortex,
\begin{equation}\label{eq:vor_vertical}
\frac{\partial F_z}{\partial z}\bigg|_{\mathrm{max}} \approx \Phi_0 A\left[\frac{1}{(z+\lambda_{ab})^2}+\frac{2h_0}{(z+\lambda_{ab})^3}\right].
\end{equation}

\begin{figure}
\centering
\includegraphics[width=1\columnwidth]{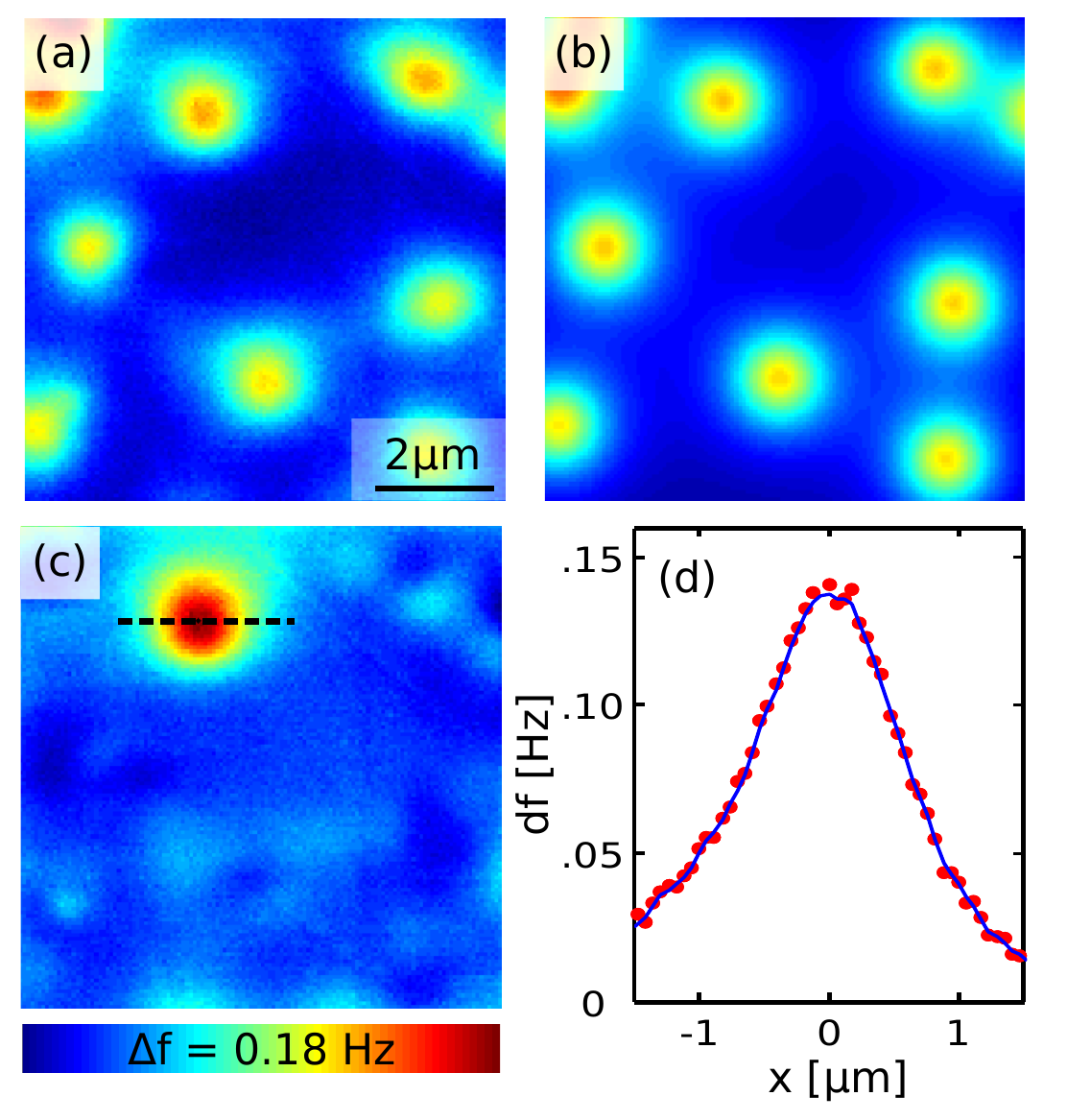}
\caption{Extraction of $(\partial F_z/\partial z)_{\mathrm{max}}$ for a single vortex. \textbf{(a)} Vortex image (raw data minus plane background) acquired at tip height $z=450$ nm. \textbf{(b)} Fit of truncated cone model in Eq.\ \ref{eq:vor_lateral} to data in (a). \textbf{(c)} Contribution of a single vortex, obtained by subtracting the fits of all-but-one vortex in (b) from the data in (a). \textbf{(d)} Cut through the single vortex in (c), along the dashed line. Red dots are data, blue line is smoothed data, and $(\partial F_z/\partial z)_{\mathrm{max}}$ is taken to be the maximum of the blue curve.}
\label{fig:fitting_proc}
\end{figure}

We first fit each image in the height series (exemplified in Figs.\ \ref{fig:PinningForce}(a-e)) to Eq.\ \ref{eq:vor_lateral}. We fix $h_0=200$ nm according to tip manufacturer specifications, and fix $z$ according to each measurement. The fit parameters are the penetration depth $\lambda_{ab}$, the tip parameter $A = \alpha t m$, and the center coordinates $(x_i,y_i)$ of each vortex. Example raw data and fit are shown in Figs.\ \ref{fig:fitting_proc}(a-b). We then subtract the fit image of all vortices but one, to leave a single vortex, exemplified in Fig.\ \ref{fig:fitting_proc}(c). This procedure effectively removes the tails of the surrounding vortices so that the remaining single vortex can be analyzed individually with higher accuracy. From this individual vortex, we read off the value of $(\partial F_z/\partial z)_{\mathrm{max}}$, as demonstrated in Fig.\ \ref{fig:fitting_proc}(d).

The values of $(\partial F_z/\partial z)_{\mathrm{max}}$ for seven vortices fully within the field of view are plotted as a function of $z$ in Fig.\ \ref{fig:PinningForce}(f). We then fit all of these points to Eq.\ \ref{eq:vor_vertical}, with fixed $h_0=200$ nm and fit parameters $A$ and $\lambda_{ab}$. This allows us to extrapolate the force gradient to infinity. We integrate this curve from $z=\infty$ to the maximum measured tip height $z_{\mathrm{max}}$, then sum the measured values of $(\partial F_z/\partial z)_{\mathrm{max}}$, to arrive at the total vertical force exerted by the vortex on the tip above the center of the vortex, shown in Fig.\ \ref{fig:PinningForce}(g).


\begin{figure}
\centering
\includegraphics[width=0.5\textwidth]{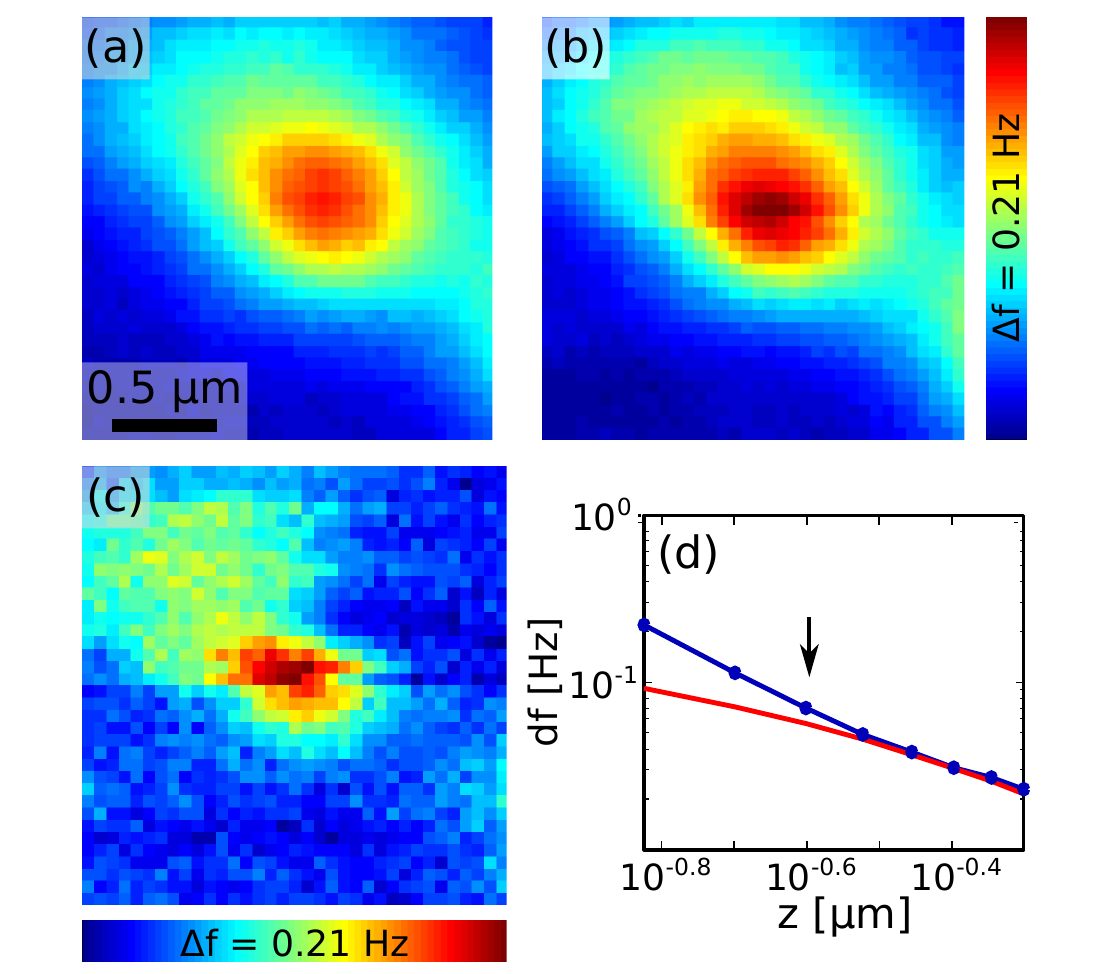}
\caption{Method to determine vortex movement. (a) A vortex image acquired at tip height $z=300$ nm. (b) The same vortex at tip height $z=250$ nm. (a) and (b) are plotted on the same color scale. (c) The difference between the vortex images in (a) and (b). (d) Log-log plot of the maximum pixel value of each successive difference image, vs.\ height $z$. Blue points are extracted from the height series data. Red line indicates the theoretical upper bound determined from Eq.\ \ref{eq:vor_lateral}, with up to one pixel lateral shift, as described in the text. Arrow shows the point corresponding to $z=250$ nm, where the slope of the data (blue) exceeds the theoretical slope (red).}
\label{fig:determine_height}
\end{figure}

To find the depinning forces, we assume that the vortex moves when the tip is scanned over the radius $R$ where the lateral force is greatest. The analytic form of the lateral force is computed by taking the lateral derivative of Eq.\ \ref{eq:tip_vortex_energy}. The ratio of $F_z^{\mathrm{max}}/F_r^{\mathrm{max}}$ obtained from the analytic form is $\sim$0.31 for $h_0$ and $z$ in our range of interest.

To determine the height at which each vortex is depinned, we look at the differences between the images taken at successive tip heights. This shows any abrupt change in vortex shape which indicates vortex movement. To make a quantitative determination, we use the vortex model in Eq.\ \ref{eq:vor_lateral} to compute the maximum theoretical value of each difference image between two successive heights, allowing also for the possibility that successive images are misaligned by one pixel to account for possible tip drift between images. We plot the measured data (blue dots) and theoretical values (red curve) on a log-log plot to highlight their difference. A measured slope of absolute value greater than the theoretical slope indicates an abrupt change larger than can be accounted for by tip height and image misalignment alone. This method is depicted in Fig.\ \ref{fig:determine_height}. Using this method, we obtain the histogram in Fig.\ \ref{fig:PinningForce}(h). By our force curve above, we obtain an average vertical force of $14.5\pm1.2$ pN at the center of the vortex, and an average maximal lateral force of $4.5\pm0.3$ pN. Including the systematic uncertainty of the cantilever spring constant, this value becomes $4.5\pm 2.5$ pN. Uncertainty in the tip truncation height $h_0$ contributes insignificantly to the total error.

We note that the depinning forces we report are an upper bound on the actual depinning forces, since we measure only at 50 nm $z$ increments, and we assume that each vortex moves where the lateral force from the tip is greatest.

\section{\label{app:PenetrationDepthDetails}Penetration Depth Analysis}

\begin{figure}
\centering
\includegraphics[width=0.75\columnwidth]{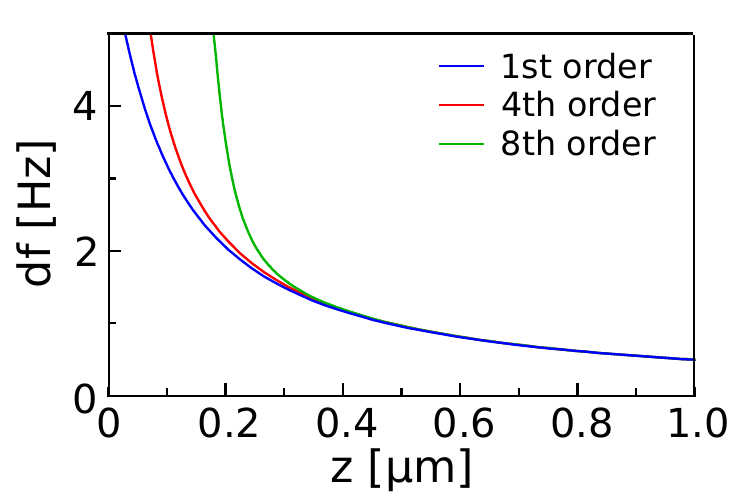}
\caption{Successive approximations of Eq.\ \ref{eq:meissner_more_terms} for the Meissner response in a truncated cone tip model. Parameters plotted here are $h_0=200$ nm, $\mu_0 A =1$ $\mu$m/Hz, and $\lambda_{ab} = 300$ nm. (We repeated the convergence test for values of $\lambda_{ab}$ up to 350 nm; in each case the series was found to converge well for $z \gtrsim \lambda_{ab}$.)}
\label{fig:meissner_approx}
\end{figure}

We model the Meissner response following Ref.\ \onlinecite{LuanThesis2011}, which we reproduce here for completeness. The Meissner response of a type-II superconductor exerts a force on an MFM tip above the $ab$ plane and far from any vortices, given by\cite{XuPRB1995}
\begin{multline}
F(z)=
\frac{\mu_0}{4\pi}\int_0^\infty dk k^3G(\lambda_{ab} k)e^{-2zk}\int_{\mathrm{tip}}d\vec{r'}\times\\
\int_{\mathrm{tip}}d\vec{r''}M(\vec{r'})M(\vec{r''})e^{-k(z'+z'')}J_0(k|\vec{R'}-\vec{R''}|)
\end{multline}
The MFM tip measures the vertical force gradient
\begin{multline}
\frac{\partial F_z}{\partial z}(z) = -\frac{\mu_0}{2\pi}\int_0^\infty dkk^4G(\lambda_{ab} k)e^{-2zk}\int_{\mathrm{tip}}d\vec{r'}\times\\
\int_{\mathrm{tip}}\vec{dr''}M(\vec{r'})M(\vec{r''})e^{-k(z'+z'')}J_0(k|\vec{R'}-\vec{R''}|)
\end{multline}
where $G(x)=(\sqrt{1+x^2}-x)/(\sqrt{1+x^2}+x)$.
Using the approximation $J_0(k|\vec{R}-\vec{R'}|)\approx J_0(0)$ and
\begin{equation}\label{eq:G approx}
G(x)=\frac{\sqrt{1+x^2}-x}{\sqrt{1+x^2}+x}\sim e^{-2x}\left(1+\frac{x^3}{3}-\frac{3x^5}{20}+O[x]^6\right)
\end{equation}
 for $x\ll 1$, we obtain, to first order $G(x)\sim e^{-2x}$,
\begin{multline}\label{eq:meissner_response}
\frac{\partial F_z}{\partial z}(z) = -2\pi \mu_0 A^2 \times\\
\left[\frac{1}{z+\lambda_{ab}}+\frac{h_0}{(z+\lambda_{ab})^2}+\frac{h_0^2}{2(z+\lambda_{ab})^3}\right]
\end{multline}
\noindent where $A=\alpha tm$ is the tip parameter.

To confirm the convergence of the series, we also compute the expression including higher order terms in the expansion of $G(x)$ from Eq.\ \ref{eq:G approx}.
\begin{widetext}
\begin{align}\label{eq:meissner_more_terms}
\frac{\partial F_z}{\partial z}(z) = -2\pi \mu_0 A^2
 &\left\{ \left[\frac{1}{(z+\lambda_{ab})}+\frac{h_0}{(z+\lambda_{ab})^2}+\frac{h_0^2}{2(z+\lambda_{ab})^3} \right] \right. \nonumber \\
 &+ \frac{\lambda_{ab}^3}{(z+\lambda_{ab})^3}\left[\frac{1}{4(z+\lambda_{ab})}+\frac{h_0}{(z+\lambda_{ab})^2}+\frac{5h_0^2}{4(z+\lambda_{ab})^3} \right] \nonumber \\
 &\left. - \frac{\lambda_{ab}^5}{(z+\lambda_{ab})^5}\left[\frac{9}{16(z+\lambda_{ab})}+\frac{27h_0}{8(z+\lambda_{ab})^2}+\frac{189h_0^2}{32(z+\lambda_{ab})^3} \right]
  +O\left[\frac{\lambda_{ab}^6}{(z+\lambda_{ab})^6}\right] \right\}.
\end{align}
\end{widetext}
\noindent We plot several partial sums for test $\lambda_{ab}$ and $h_0$ values, exemplified in Fig.\ \ref{fig:meissner_approx}, which shows that the series converges rapidly in the range $z\gtrsim \lambda_{ab}$, while it does not converge well for $z\ll \lambda_{ab}$. Thus we are justified in using the first order approximation (Eq.\ \ref{eq:meissner_response}) for $z \gtrsim \lambda_{ab}$.


\begin{figure*}[ht]
\center
\includegraphics[width=1.5\columnwidth]{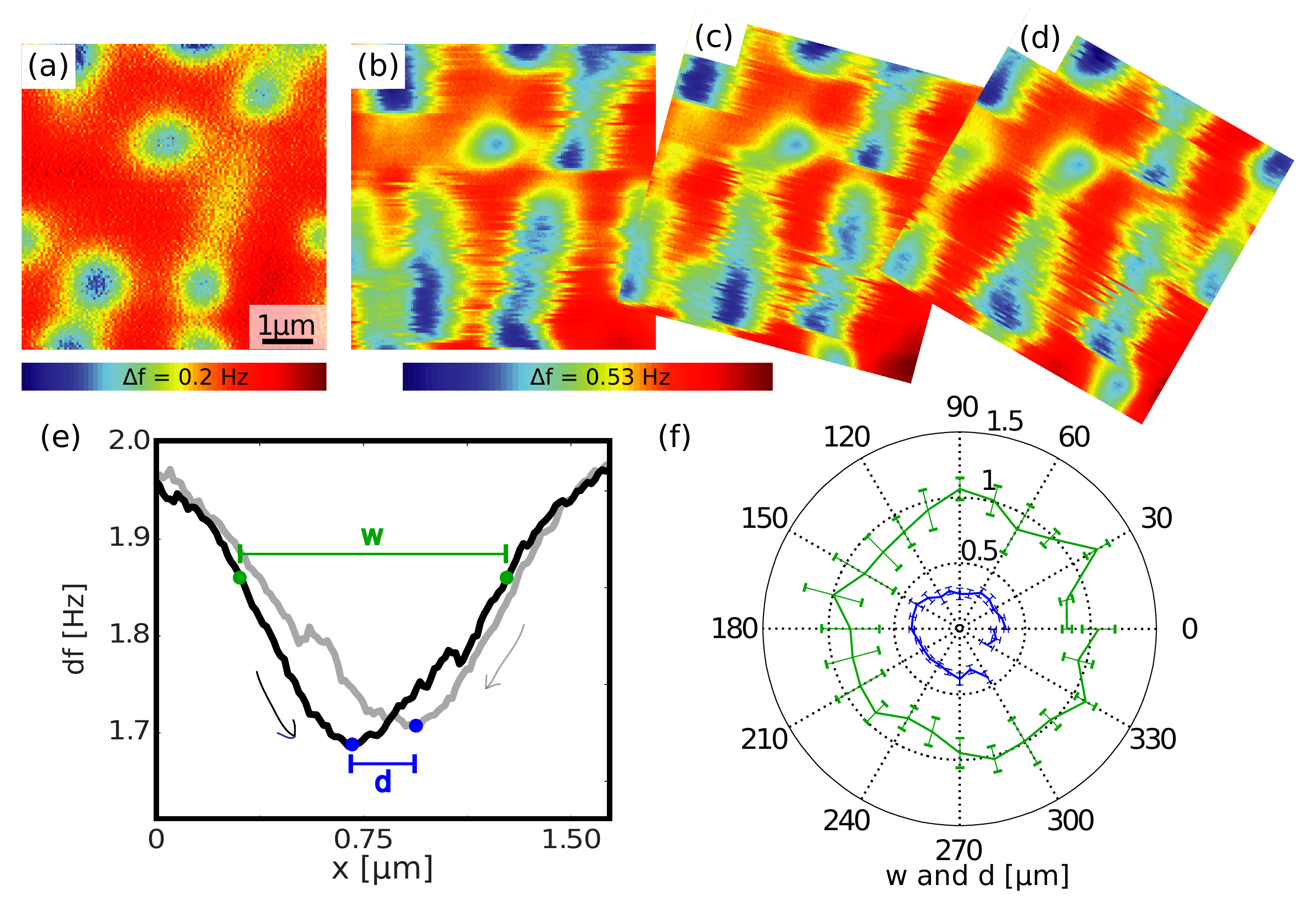}
\caption{Vortex wiggling anisotropy. (a-d) Series of four rotated images exemplifying the data acquisition. Twenty-four such images were acquired over 360$^{\circ}$ with $15^{\circ}$ resolution. (e) Two metrics for vortex displacement: (i) Width at 30\% of the vortex height, corresponding approximately to the radius of maximum lateral force between tip and vortex, shown in green; (ii) Displacement between the peaks of the forward and backward scans, shown in blue. (f) Polar plot showing no significant anisotropy using either displacement metric. Polar plot data is shown for a single vortex, but is representative of similar plots for 5 different vortices.}
\label{fig:VortexWigglingAnisotropy}
\end{figure*}

To measure the Meissner response curve, we scan the tip from high to low above the sample at a fixed location $\sim$2 $\mu$m laterally from the nearest vortex, as plotted in Fig.\ \ref{fig:MeissnerLambda}(a). We see that the force curve changes abruptly for $z<100$ nm. We attribute this feature to electrostatic forces between the tip and the sample. However, this force gradient falls off as $1/z^3$, so we may neglect it for our tip heights of interest $z > 200$ nm, where the dominant Meissner force gradients fall off as $1/z$.

We fit the Meissner response curve in Fig.\ \ref{fig:MeissnerLambda}(a) from a cutoff height of $z_{\mathrm{cut}}=300$ nm to Eq.\ \ref{eq:meissner_response}, with fixed $h_0$, and fit parameters $A$ and $\lambda_{ab}$. We find $\lambda_{ab} = 320 \pm 60$ nm. The greatest uncertainty comes from our uncertainty in the tip truncation height $h_0$, which is nominally 200 nm according to manufacturer specifications, but which we have not measured directly. Thus the error bars are determined by redoing the fit with $h_0=100$ nm and $h_0=300$ nm. We note that our choice of cutoff height $z_{\mathrm{cut}}$ within the range 200-300 nm does not change the final result.

Alternatively, we may obtain Meissner response curves from the height series of vortex images in Figs.\ \ref{fig:PinningForce}(a-e). For each image that we fit, there is a constant offset as a fit parameter. We attribute this constant offset to the Meissner response of the superconductor to the tip. Plotting this constant offset with respect to the image tip height gives the equivalent of a Meissner response curve. Fitting the same model, Eq.\ \ref{eq:meissner_response}, to the data points obtained this way gives a consistent value of $\lambda_{ab} = 300\pm 60$ nm, as shown in Fig.\ \ref{fig:MeissnerLambda}(b).

Our MFM measurements allow us to extract $\lambda_{ab}$ in several independent ways, all assuming a truncated cone tip shape. First, our fits of Eq.\ \ref{eq:vor_lateral} to the vortex images exemplified in Figs.\ \ref{fig:PinningForce} gave $\lambda_{ab}\sim500$ nm. However, the finite lateral extent of the tip typically leads to an over-estimation of $\lambda_{ab}$ using this kind of analysis.\cite{LuanThesis2011} Second, the best fit of the $(dF_z/dz)_{\mathrm{max}}$ data in Fig.\ \ref{fig:PinningForce}(f) to Eq.\ \ref{eq:tipVortex} gives shows $\lambda_{ab}=380\pm50$ nm. This method is typically less sensitive to the details of tip shape.\cite{LuanThesis2011} Finally, the Meissner analysis presented here in Appendix \ref{app:PenetrationDepthDetails} depends only on data acquired far from vortices, where there is no lateral spatial dependence, and consequently very little sensitivity to the tip shape. We therefore find the Meissner method to be the most reliable measure of $\lambda_{ab}$

\section{\label{app:Anisotropy}In-Plane Anisotropy}

In-plane anisotropy in Fe-SCs can arise from anisotropy of the crystal structure,\cite{ChuScience2010, TanatarPRB2010} defects,\cite{SongPRL2012, AllanNatPhys2013} band structure,\cite{WangPRB2012} or superconducting gap.\cite{SongScience2011}

\vspace{1mm} \noindent \textbf{Structural Anisotropy}\\
Early phase diagrams for the `1111' family of Fe-SCs, including La-1111 (Refs.\ \onlinecite{HuangPRB2008, LuetkensNatMat2009, HessEPL2009}), Ce-1111 (Ref.\ \onlinecite{ZhaoArxiv2008}), Pr-1111 (Ref.\ \onlinecite{RotunduPRB2009}), Nd-1111 (Ref.\ \onlinecite{MalavasiJACS2010}), and Sm-1111 (Refs. \onlinecite{HessEPL2009, KamiharaNJP2010}) show no overlap between orthorhombic structure and superconductivity. However, in most families of Fe-SCs, the orthorhombic phase overlaps with the superconducting phase,\cite{StewartRMP2011} and several more recent papers have questioned the non-overlap in Ce-1111 (Ref.\ \onlinecite{ZhaoNatMat2008}) and Sm-1111 (Refs.\ \onlinecite{MargadonnaPRB2009, MartinelliPRL2011}), suggesting that weak orthorhombicity may persist nanoscale domains, making it hard to detect using standard procedures. The possible coexistence of orthorhombic phase and superconductivity in `1111' materials remains controversial.\cite{RotunduPRL2013, MartinelliPRL2013} Our own EBSD data in Fig.\ \ref{fig:OrientationTc}(d) cannot distinguish between orthorhombic and tetragonal state in \Nd.

\vspace{1mm} \noindent \textbf{Electronic Anisotropy}\\
Electronic anisotropy could arise from the underlying band structure or the superconducting gap. Theoretical calculations predict negligible anisotropy of the bulk Fermi surface of \Nd\ (Ref.\ \onlinecite{LiuPRB2010}). This prediction is apparently born out in ARPES measurements,\cite{LiuPhysicaC2009} although one must be cautious because the ARPES-measured Fermi surface is significantly larger than would be expected for the bulk bands, and may not be representative of the bulk. ARPES also measured negligible superconducting gap anisotropy on \Nd (Ref. \onlinecite{KondoPRL2008}), although again one must caution that ARPES may be sensitive only to surface superconductivity which may not be representative of the bulk. There is some evidence that at least some Fe-SCs do have significant superconducting gap anisotropy, such as the $d$-wave KFe$_2$As$_2$ (Ref.\ \onlinecite{ReidPRL2012}), but $T$-dependence of $\lambda$ in \Nd\ seems to rule out the $d$-wave gap symmetry.\cite{MartinPRL2009}

\vspace{1mm} \noindent \textbf{Defect Anisotropy}\\
Fe site defects in Fe-SCs have shown anisotropy on two length scales. First, in both tetragonal and orthorhombic structural phases, Fe site defects form simple geometric dimers of two-unit-cell extent, randomly oriented along either of the two orthogonal Fe-As directions. These structures have been observed by STM in FeSe (Ref.\ \onlinecite{SongPRL2012}), LiFeAs (Refs.\ \onlinecite{HanaguriPRB2012, GrothePRB2012}), and LaFeAsO (Ref.\ \onlinecite{ZhouPRL2011}). Second, in the spin density wave phase, Fe site defects form electronic dimers of size $\sim$8 to $\sim$16$a_{\mathrm{Fe-Fe}}$. These dimers are uniformly oriented along the $a$ (longer, antiferromagnetic) axis. They have been observed directly in FeSe (Ref.\ \onlinecite{SongPRL2012}) and NaFeAs (Ref.\ \onlinecite{RosenthalNatPhys2013}) and inferred indirectly in Ca(Fe$_{1-x}$Co$_x$)$_2$As$_2$ (Ref.\ \onlinecite{AllanNatPhys2013}). Their appearance is explained by a competition between the impurity-induced $(\pi,\pi)$ magnetism and the surrounding bulk spin density wave $(\pi,0)$ phase.\cite{GastiasoroPRB2014}

\vspace{1mm} \noindent \textbf{Search for Anisotropy in Vortex Wiggling}\\
To search for signatures of in-plane anisotropy, we investigate the angle-dependence of vortex wiggling. This technique has been used to reveal pinning anisotropy in YBa$_2$Cu$_3$O$_{7-x}$ (Ref.\ \onlinecite{AuslaenderNatPhys2008}). Here we similarly raster the tip across a set of vortices at a series of angles, as illustrated in Figs.\ \ref{fig:VortexWigglingAnisotropy}(a-d). The average wiggle distance of a vortex at each angle can be quantified in two different ways, as illustrated in Figs.\ \ref{fig:VortexWigglingAnisotropy}(e-f). We found no statistically significant anisotropy of vortex wiggling using either metric, on 5 different vortices.


\begin{thebibliography}{79}
\bibitem [{can()}]{cantilever}%
  \BibitemOpen
  \href@noop {} {}\bibinfo {note} {As of 2013, the current NSC18 cantilevers
  from Mikromasch are specified with $k_{\mathrm{typ}} = 2.8$ N/m
  ($k_{\mathrm{min}} = 1.2$ N/m and $k_{\mathrm{max}} = 5.5$ N/m. An old NSC18
  cantilever from 2008 was used for this experiment, with
  manufacturer-specified $k=3.5\pm 1.5$ N/m.}\BibitemShut {Stop}%
\end{thebibliography}

%

\end{document}